\documentclass[preprint,aps,showpacs]{revtex4}
\DeclareMathAlphabet{\mathpzc}{OT1}{pzc}{m}{it}
\usepackage{tipa}
\usepackage{mathrsfs}
\usepackage{tabularx}
\usepackage{dcolumn}
\newcolumntype{C}{>{\centering\arraybackslash}X}
\newcolumntype{R}{>{\raggedleft\arraybackslash}X}
\newcolumntype{L}{>{\raggedright\arraybackslash}X}
\usepackage{graphics}
\usepackage{dcolumn}
\usepackage{bm}
\usepackage[dvips]{graphicx}
\usepackage[dvips]{rotating}
\usepackage[latin1]{inputenc}
\usepackage{dcolumn}
\usepackage{tabularx}
\usepackage{amsmath}
\usepackage{amssymb}
\usepackage{footnote}

\begin{document}
\draft
\title{Circular Rydberg states of atomic hydrogen in an arbitrary magnetic field}
\author{L. B. Zhao$^{\dag}$, B. C. Saha$^{\dag}$, and M. L. Du$^{\S}$,}
\affiliation{$^{\dag}$Department of Physics, Florida A $\&$ M University, Tallahassee, Florida 32307, USA}
\affiliation{$^{\S}$Institute of Theoretical Physics, Chinese Academy of Sciences, P. O. Box 2735,
Beijing 100190, China}
\begin{abstract}
We report a theoretical scheme using a B-spline basis set to improve the poor computational
accuracy of circular Rydberg states of hydrogen atoms in the intermediate magnetic field.
This scheme can produce high accuracy energy levels and valid for an arbitrary magnetic field.
Energy levels of hydrogen are presented for circular Rydberg states with azimuthal quantum
numbers $|m|$ = 10 - 70 as a function of magnetic field strengths ranging from zero to 2.35
$\times$ 10$^9$ T. The variation of spatial distributions of electron probability densities with
magnetic field strengths is discussed and competition between Coulomb and magnetic
interactions is illustrated.
\end{abstract}
\date{\today}
\maketitle
\section{introduction}
Circular states of Rydberg atoms are states with $|m|$ = $n$ - 1, where $m$ and $n$
are the azimuthal and principal quantum numbers, respectively. The word ``circular"
here is used because the classical orbits corresponding to these quantum states are
circular. Such states have drawn considerable interests since the early development
stage of quantum theory. In modern physics, circular Rydberg states play a crucial role
in understanding the correspondence limit between classical and quantum physics.

Rydberg atoms in circular states possess some special properties, such as the largest
magnetic moment, the smallest Stark effect, the longest radiative lifetime, and highly
anisotropic collision cross sections \cite{Hulet83,Lutwak97}. These properties make
them applicable in many research domains, including cavity QED \cite{Gleyzes07},
quantum computation and information processing \cite{Brune96,Hagley97},
measurements of fundamental physics constants \cite{Liang86}, ultrafast dynamics
of Rydberg atoms \cite{deBoer93}, and so on. The earliest experiment for circular
state populations was made by Hulet and Kleppner \cite{Hulet83}. In that
experiment, the circular states of lithium atoms with $|m|$ = 17, 18 were prepared
using an adiabatic microwave transfer technique. Soon afterwards, the same technique
was applied to the other two alkali metal atoms, cesium \cite{Hulet85} and rubidium
\cite{Nussenzveig94}. The external electric and magnetic fields have been adopted
to control adiabatic microwave transfer and to split $\pm$$m$ degeneracy,
respectively \cite{Nussenzveig94,Nussenzveig91}. A theoretical scheme using
crossed electric and magnetic fields to populate circular states was proposed by
Delende and Gay \cite{Delande88} and realized experimentally by Hare et al.
\cite{Hare88}. The creation of circular states of hydrogen atoms was based on
both adiabatic microwave transfer and the crossed fields method \cite{Lutwak97}.
Very recently, Maeda et al. \cite{Maeda09} successfully made long-lived,
nondispersing Bohr wave packets starting from Li Rydberg atoms in circular states.
Obviously interests in circular states have been enhanced in recent years.

Besides fundemental physics interests, investigation of magnetized H atoms is of
application importance in astronomy and astrophysics. To understand the behavior
of magnetized white dwarf stars (10$^2$ $-$ 10$^5$ T) and neutron stars
(10$^7$ $-$ 10$^9$ T), the accurate knowledge of the physical properties of
atoms in strong magnetic fields is essential.

Much attention has been paid to experiment and astronomical observation, but
corresponding computations of
circular Rydberg states of atoms in external fields are relatively rare. Wunner
et al. \cite{Wunner86} solved the Sch\"odinger equations for circular states
of magnetized hydrogen atoms. For low fields, they expanded wave functions in
terms of a complete, orthonormal set of functions in the radial direction and
spherical harmonics in the angular direction, while for high fields ($\le$ 10$^7$ T),
they employed the adiabatic approximation. In the intermediate field range, they
smoothly jointed the two-side results. Seemingly, the method of Wunner et al.
may effectively handle any magnetic field from low to high. However, that is
not true. It is widely known that the adiabatic approximation is valid only
for extremely high magnetic fields, and interpolation in the intermediate field
range may produce poor results because of the used energies stemming from the
adiabatic approximation. Here it is necessary to point out that {\it intermediate
magnetic field} used in literature is a very ambiguous concept. Strictly speaking in
physics, the ``intermediate magnetic field" means a magnetic field when Coulomb
and magnetic interactions are of a comparable magnitude. In other words, also
it is closely relevant to the energy levels besides magnetic fields.

Liu et al. \cite{Liu93} performed calculations of circular Rydberg states of H atoms
in magnetic field strengths ranging from zero to 70.5 T, based on a B-spline expansion
method, in which wave functions are expanded in terms of B-spline functions in
both the radial and angular directions. However, their method is limited to low
magnetic fields. It should be emphasized, again, that the ambiguous concept of
the so-called low magnetic fields is being used here. The disadvantages of their
method have been analyzed recently in Ref. \cite{Zhao07}. The accuracy of their
energy levels decreases with magnetic fields $B$ when $B>$ 47 T.
Germann et al. \cite{Germann95} developed a dimensional perturbation
theory to evaluate circular Rydberg states of magnetized atomic hydrogen with
an extreme high accuracy. The theory is applicable to entire range of magnetic
field strengths, but is limited to high azimuthal quantum number ($m\gg$ 1).
Later, Watson and coworker \cite{Walkup01} extended their method to include
low-$m$ situations. However, not in all field strengths, circular states may be
treated because of the limitation of this theory.

This paper focuses on the problem of circular Rydberg states in the intermediate
magnetic fields. First of all, we develop a method to evaluate circular Rydberg
states of atomic hydrogen in a strong magnetic field. In cylindrical coordinates,
the wave functions are expanded in terms of Landau states in the $\rho$ direction
and the B-spline basis in the $z$ direction. This method is combined together with a
recently reported finite-basis-set technique \cite{Zhao07}, which is valid for weak
fields. We will show that the combination scheme is a high accuracy tool to
compute of circular Rydberg states of atomic hydrogen in the intermediate magnetic
fields, and valid for an arbitrary magnetic field.

\section{theoretical method}
The nonrelativistic Hamiltonian of hydrogen atoms on a uniform magnetic
field along the $z$ axis reads in atomic units
\begin{equation}
{\widehat H}=-\frac 12\nabla ^2-\frac 1{\sqrt{\rho^2+z^2}} +
\frac{\gamma}2({\hat \ell}_z + 2{\hat s}_z) + \frac18{\gamma^2}\rho^2,
\label{Hami}
\end{equation}
where $\gamma$ is the magnetic field strength in units of
$B_0 \simeq 2.35\times10^5$ T, ${\hat \ell}_z$ and ${\hat s}_z$ are the
$z$ components of the orbital and spin angular momenta, respectively, the
third term linear in $\gamma$ is the paramagnetic potential, and the fourth
term quadratic in $\gamma$ is the diamagnetic potential. It is easily seen
that ${\widehat H}$ commutes with ${\hat \ell}_z$ (corresponding quantum
number $m$) as well as ${\hat s}_z$. Depending on relative magnitudes of
Coulomb and diamagnetic potentials, the symmetry of the system changes.
The system is more spherically symmetric if the Coulomb potential is dominant,
and more cylindrically symmetric if the diamagnetic potential is dominant.

For strong magnetic fields, we solve the Schr\"odinger equation in cylindrical
coordinates by expanding wave functions in terms of Landau states in the
$\rho$ direction and the B-spline basis in the $z$ direction. The wave function
is expressed as
\begin{equation}
\Psi (\rho,z,\phi)=\sum_{i,n}C_{in}B_{i}(z){\mathcal R}_n(\rho,\phi),
\label{WF}
\end{equation}
where ${\mathcal R}_n(\rho,\phi)$ is the normalized wave function for the Landau
state with $n\ge0$, given by \cite{Landau}
\begin{equation}
{\mathcal R}_n(\rho,\phi)={\mathscr R}_n(\rho)\frac{e^{im\phi}}{\sqrt{2\pi}} =
\frac{\gamma^{(|m|+1)/2}}{m!}
\sqrt{\frac{(|m|+n)!}{2^{|m|}n!}}e^{-\rho^2\gamma/4}\rho^{|m|}
{_1F}_1(-n,|m|+1,\rho^2\gamma/2)\frac{e^{im\phi}}{\sqrt{2\pi}},
\label{Landaustate}
\end{equation}
with ${_1F}_1$ being the confluent hypergeometric function, and $B_{i}(z)$ are the
B-spline functions. The B-spline functions along with their properties have been
outlined in Ref. \cite{Zhao07}, and a complete description can be also found in the
book of de Boor \cite{deBoor}. Therefore, here it is unnecessary to give any elaboration
once again. ${\mathcal R}_{n}(\rho,\phi)$ in Eq. (\ref{Landaustate}) should have
been denoted as ${\mathcal R}_n^m(\rho,\phi)$, but the superscript $m$ is dropped
for simplification. Considering $m$ is a good quantum number, doing so should not
give rise to any confusion. Substituting Eq. (\ref{WF}) into the Schr\"odinger equation
with the Hamiltonian (\ref{Hami}), and projecting onto the basis $B_{i}{\mathcal R}_{n}$
yield the following matrix equation
\begin{equation}
HC=E\mathcal{N}C,
\label{matrix}
\end{equation}
where the Hamiltonian matrix elements are of the form
$$
H_{i'n';in}=-\frac{1}{2}\delta_{n'n}\int_0^{z_{max}}B_{i'}(z)
\frac{d^2}{dz^2}B_{i}(z)dz +
n\gamma\delta_{n'n}\int_0^{z_{max}}B_{i'}(z)B_{i}(z)dz
$$
\begin{equation}
+\int_0^{z_{max}}B_{i'}(z)V_{n'n}(z)B_{i}(z)dz,
\label{Hmatrixelement}
\end{equation}
and $\mathcal{N}$ is the overlap matrix with elements
\begin{equation}
{\mathcal N}_{i'n';in} = \delta_{n'n}\int_0^{z_{max}}
B_{i'}(z)B_{i}(z)dz.
\label{mateleN}
\end{equation}
The upper limits of the integration, $z_{max}$, in Eqs. (\ref{Hmatrixelement},\ref{mateleN})
denotes the maximum of coordinate $z$. It changes with energy levels and magnetic field
strengths. $z_{max}$ should be taken to be large enough to obtain stable eigenvalues in
calculations. The overlap matrix $\mathcal{N}$ is a positive definite real symmetric matrix.
Its existence is attributed to the B-spline basis nonorthogonality. The matrix element
$V_{n'n}(z)$ in Eq. (\ref{Hmatrixelement}) is given by \cite{Friedrich83}
$$
V_{n'n}(z)= -\int_0^{\infty} \mathscr{R}_{n'}(\rho)\frac{1}{\sqrt{\rho^2+z^2}}
\mathscr{R}_n(\rho)\rho d\rho $$
\begin{equation}
= -\frac2{\sqrt{\pi}}\sum_{j = 0}^{\min(n',n)}
\frac{[n'!n!(n'+|m|)!(n+|m|)!]^{1/2}}{(n'-j)!(n-j)!(|m|+j)!j!}\mathcal{I}(n'+n-2j,2j+|m|;z),
\label{Vn'n}
\end{equation}
with
\begin{equation}
\mathcal{I}(\alpha,\beta;z)=\sqrt{\frac{\gamma}2}\int_0^{\pi/2}(\sin x)^{2\beta}
(\cos x)^{2\alpha}\exp{\left[-\frac{\gamma z^2 {\cos^2x}}{2{\sin^2x}}\right]}dx.
\label{mathcalI}
\end{equation}
The adaptive Gauss-Kronrod quadrature and the Gaussian quadrature are, respectively,
used to evaluate the integral of Eq. (\ref{mathcalI}) numerically, and all the matrix
elements of $H$ and $\mathcal{N}$. For states of even and odd parity, we perform
calculations at the whole space of $z$ with $-z_{max}\le z \le z_{max}$ and the
half space of $z$ with $0 \le z \le z_{max}$, respectively. The combination of such a
choice and properties of B-spline functions make it easy to enforce boundary conditions.
That is to simply remove $i=1,N$ terms from the summation over $i$ in Eq. (\ref{WF}),
namely $B_1(z)$ and $B_N(z)$, where $N$ represents the number of B-spline functions.
Once the calculations of the matrix elements are finished, standard routines for matrix
diagonalization can be used to obtain the eigenvalues and eigenvectors numerically.

It should be pointed out that this method is valid only for strong fields, because the
basis size required to obtain stable convergence becomes larger and large with
decreasing field strengths. In the situation of weak fields, we will adopt the
finite-basis-set technique \cite{Zhao07}, in which the Schr\"odinger equation is
solved in spherical coordinates by expanding wave functions in terms of a B-spline
basis in the radial direction and spherical harmonics in the angular direction. The
finite-basis-set technique was developed recently in order to calculate ground and
lowly excited states for magnetized hydrogen atoms. Its details have been presented
and therefore are omitted here. We will combine the current method and the
finite-basis-set technique for computations of circular Rydberg states of atomic
hydrogen for arbitrary magnetic field strengths.

\section{Results and discussion}
Circular Rydberg states with $|m| = 10 - 70$ were calculated in a wide field strength
range. Only the eigenvalue problem for states with negative $m$ is solved, since
eigenvalues for states with positive $m$ are easily obtained by an energy shift from
the corresponding states of negative $m$ \cite{Wunner86}. The maximum number
and order of B-spline functions were taken to be $N=100$ and $k=11$, respectively.
A crucial procedure in the current calculation is to suitably select the knot sequence
of B-spline functions and maxima of space coordinates, $z_{max}$ and $r_{max}$,
where $r_{max}$ represents the maximum of radial coordinates $r$. For
calculations in cylindrical coordinates with the current method, $z=\{0,z_{max}\}$
was divided into the inner and outer region, and in each region an equidistant mesh
was taken. Furthermore, the mesh size in the inner region was selected to be equal
to twice the mesh size in the outer region. Such a choice is expected to characterize
the behavior of circular electron motion in the combined Coulomb and magnetic
fields quite well. For calculations in spherical coordinates with the finite-basis-set
technique, an exponential mesh was adopted, following Ref. \cite{Zhao07} (see this
reference for details). Energy levels were optimized by adjusting the distribution of
the knots, and $z_{max}$ (or $r_{max}$).

For any given state, there does exist a magnetic field range with strong competition
between Coulomb and magnetic interactions. In this range, the solution of the
Schr\"odinger equation is very difficult. Either a gap of fields, where energy levels
can not be directly calculated because of the limitation of the used method, remain
left or the gap is smoothly jointed to the results from the two sides of the gap (see,
i.e., Ref. \cite{Wunner86} and references therein). Our method, however, is very
effective in such a field range. Table I lists eigenvalues for circular states with
$m=-$ 10 and $-$ 70 in a magnetic field with strengths ranging from 0.0001 to
0.1 a.u., calculated in the current method and the finite-basis-set technique. The
uncertainty of each eigenvalue is $\pm$ 1 in the last digit, except for those in a
range from 0.0001 to 0.002 a.u. in the second column and from 0.02 to 0.1 a.u.
in the third column. The two methods are in excellent agreement.

The numbers of spherical harmonics and Landau states, which are used to expand
wave functions, are also given in this table. The numbers reflect symmetry of the
system at a given state and field strength because of the competition between
Coulomb and magnetic interactions. For example, let us look at the circular state
with $m=-$ 10. At $\gamma=0.0001$ a.u., the system is more spherically
symmetric (the number of spherical harmonics is 2), as the Coulomb interaction
is dominant, while the magnetic interaction is relatively small, but at
$\gamma=0.1$ a.u., the symmetry is completely broken (the number of spherical
harmonics and Landau states is close), as the two kinds of interactions become
comparable. Now, let us turn to the $m=-$ 70 state. At $\gamma=0.1$ a.u., the
system is more cylindrically symmetric (the number of Landau states is 5), as the
magnetic interaction is dominant, while the Coulomb interaction is relatively small,
but at $\gamma=0.0001$ a.u., the symmetry is completely broken, because the
two interactions become comparable.

It is time saving to adopt the different method for systems with more spherical
and more cylindrical symmetries. Totally, the method developed in this paper is
not suitable for weak magnetic fields, while the finite-basis-set technique is not
suitable for strong fields. However, both methods remain valid for intermediate
fields, and should be in excellent agreement. Eigenvalues are calculated and
listed in Table II for circular Rydberg states with $m$ from $-$ 20 to $-$ 60 in
a magnetic field range from zero to 10$^4$ a.u. We first used the finite-basis-set
technique when the magnetic field is small and then switched to the current
method when it becomes larger than some value.

We made a comparison between the present eigenvalues for the circular state
with $m=-$ 24 with available theoretical data \cite{Wunner86,Liu93,Germann95},
and excellent agreement is found. The compared data were not listed here as
doing so is obviously superfluous. Restricted by the used spherical coordinate
system, the method of Liu et al. \cite{Liu93} is ineffective for higher fields. One
may see that their accuracy decreases with increasing field strengths. At
$\gamma=$ 0.0003 a.u., our energy agrees very well with the highly accurate
result of Germann et al. \cite{Germann95}. Our calculation gives
$-$5.315 899 713 035 a.u. As previously mentioned, here we point out again that
not in all field strengths circular states may be calculated with the dimensional
perturbation theory of Germann et al. because of the limitation of this theory
(also see Ref. \cite{Walkup01}).

Contour plots of probability density distributions of the electron in a circular state
with $m=-$ 24 are drawn in Fig. 1 at several field strengths. This figure illustrates
the change of probability density distributions of the circular Rydberg electron with
magnetic fields. It is seen that the electron cloud becomes squeezed toward the
$z$-axis with increasing magnetic fields. At the field-free case, the probability
density widely distributes near $\rho=$ 620 a.u., but the electron cloud shrinks
in a very narrow region near $\rho=$ 1 a.u. when the field strength increases up
to $\gamma=$ 100 a.u. Furthermore, it is shown that the peak value of probability
densities at $\gamma=$ 100 a.u. is 5 orders of magnitude larger than that at
$\gamma=$ 0 a.u. (see the color bars of the figure).

To illustrate competition between Coulomb and magnetic interactions, probability
density distributions ${\mathcal P(\rho,z)}$ of the electron in a non-circular state
$n\ell_m=7g_{-4}$ are drawn in Fig. 2 at several magnetic field strengths. The
readers are referred to Ref. \cite{Zhao07} for the definition of state $n\ell_m$.
Also we calculated ${\mathcal P(\rho,z)}$ using the analytical formula of wave
functions for the field-free case and found excellent agreement with the current
numerical results. At $\gamma=$ 0.003 a.u., the figure shows the system is
more spherically symmetric. Only six spherical partial waves are needed to obtain
the energy level convergent up to 10 digits of precision. From 0.006 a.u., the
outermost and intermediate electron clouds are separated because the magnetic
interaction is becoming significant. At $\gamma=$ 0.1, the Lorenz force is
dominant. To reach the same convergence as 16 Landau states, we contained
the 38 spherical partial waves.

\section{Summary and Conclusion}
A theoretical scheme using a B-spline basis set has been proposed to compute
circular Rydberg states of hydrogen atoms in the intermediate magnetic field.
We solved the Schr\"odinger equation in the spherical and cylindrical coordinate,
corresponding to weak and strong magnetic fields, respectively. The scheme is
valid for an arbitrary magnetic field, and can produce high accuracy energy levels
(Note: the high accuracy here indicates that in the theoretical framework defined).
We showed how to save computational time as well as to obtain a high accuracy.
The energy levels of atomic hydrogen are presented for circular Rydberg states
with azimuthal quantum numbers $|m|$ = 10 - 70 as a function of magnetic field
strengths ranging from zero to 2.35 $\times$ 10$^9$ T.  The competition between
Coulomb and magnetic interactions is discussed and illustrated.
\acknowledgments
We acknowledge financial support from NSF grant No. 0630370 and NSFC grant
Nos. 90403028 and 11074260.

\begin{sidewaystable*}
\begin{center}
\caption{Eigenvalues for circular Rydberg states with $m=-$ 10 and $-$ 70
in units of Hartree in an intermediate magnetic field strength, calculated with
the finite-basis-set technique (Sph.) and the method developed in this paper (Lan.). }
\vspace{2mm}
\begin{tabular}{llllllllllll}
\toprule
&& \multicolumn{4}{c}{$m$ = $-$ 10} && \multicolumn{4}{c}{$m$ = $-$ 70} \\
\cline{3-6}\cline{8-11}
\multicolumn{1}{c} {$\gamma$ (a.u.)} &&
    \multicolumn{1}{c}{Sph.} &&
    \multicolumn{2}{c}{Lan.} &&
    \multicolumn{2}{c}{Sph.} &&
    \multicolumn{1}{c}{Lan.} \\
\colrule
1.0[$-$4]$^a$ &&$-$4.662 371 999[$-3$] (2)$^b$ &&$-$4.3[$-3$] &(42)$^c$
     &&$-$8.572 099 923[$-4$] &(12)$^b$&&$-$8.572 099 923[$-$4] (14)$^c$ \\
2.0[$-$4] &&$-$5.153 983 634[$-3$] (3) &&$-$5.11[$-3$] &(42)&&$-$1.197 805 639[$-3$] &(13)&&$-$1.197 805 639[$-$3] (10)  \\
4.0[$-$4] &&$-$6.034 883 843[$-3$] (3) &&$-$6.033[$-3$] &(42)&&$-$1.677 735 849[$-3$] &(15)&&$-$1.677 735 849[$-$3] (9)  \\
6.0[$-$4] &&$-$6.805 811 739[$-3$] (4) &&$-$6.805 7[$-3$] &(42)&&$-$2.044 675 106[$-3$] &(17)&&$-$2.044 675 106[$-$3] (8)  \\
8.0[$-$4] &&$-$7.493 913 635[$-3$] (4) &&$-$7.493 90[$-3$] &(42)&&$-$2.353 193 313[$-3$] &(19)&&$-$2.353 193 313[$-3$] (8)  \\
1.0[$-$3] &&$-$8.118 682 764[$-$3] (5) &&$-$8.118 677 [$-$3] &(42)&&$-$2.624 416 300[$-$3] &(20)&&$-$2.624 416 300[$-$3] (8)  \\
2.0[$-$3] &&$-$1.065 347 079[$-$2] (6) &&$-$1.065 347 071[$-$2] &(42)&&$-$3.683 746 040[$-$3] &(25)&&$-$3.683 746 040[$-$3] (8)\\
4.0[$-$3] &&$-$1.431 074 262[$-$2] (8) &&$-$1.431 074 262[$-$2] &(42)&&$-$5.170 506 039[$-$3] &(31)&&$-$5.170 506 039[$-$3] (7) \\
6.0[$-$3] &&$-$1.712 121 547[$-$2] (8) &&$-$1.712 121 547[$-$2] &(40)&&$-$6.303 586 798[$-$3] &(33) &&$-$6.303 586 798[$-$3] (7)\\
8.0[$-$3] &&$-$1.948 330 271[$-$2] (9) &&$-$1.948 330 271[$-$2] &(36)&&$-$7.254 178 968[$-$3] &(37) &&$-$7.254 178 968[$-$3] (7) \\
1.0[$-$2] &&$-$2.155 675 692[$-$2] (10)&&$-$2.155 675 692[$-$2] &(31)&&$-$8.088 428 422[$-$3] &(41) &&$-$8.088 428 422[$-$3] (7) \\
2.0[$-$2] &&$-$2.961 494 612[$-$2] (13)&&$-$2.961 494 612[$-$2] &(29)&&$-$1.133 550 878[$-$2] &(42) &&$-$1.133 550 878[$-$2] (6)\\
4.0[$-$2] &&$-$4.080 666 269[$-$2] (15)&&$-$4.080 666 269[$-$2] &(26)&&$-$1.586 702 29[$-$2] &(42) &&$-$1.586 702 312[$-$2] (5)\\
6.0[$-$2] &&$-$4.924 844 850[$-$2] (18)&&$-$4.924 844 850[$-$2] &(23)&&$-$1.930 346 3[$-$2] &(42) &&$-$1.930 346 459[$-$2] (5)\\
8.0[$-$2] &&$-$5.627 721 061[$-$2] (19)&&$-$5.627 721 061[$-$2] &(20)&&$-$2.217 639 3[$-$2] &(42) &&$-$2.217 639 819[$-$2] (5)\\
1.0[$-$1] &&$-$6.240 840 422[$-$2] (21)&&$-$6.240 840 422[$-$2] &(20)&&$-$2.469 076[$-$2] &(42) &&$-$2.469 077 604[$-$2] (5)\\
\botrule
\multicolumn{11}{l}{$^aA[B] = A \times 10^{B}$}\\
\multicolumn{11}{l}{$^b$The number in parenthesis gives the number of spherical harmonics.}\\
\multicolumn{11}{l}{$^c$The number in parenthesis gives the number of Landau states.}
\end{tabular}
\end{center}
\label{circularI}
\end{sidewaystable*}
\begin{sidewaystable*}
\begin{center}
\caption{Eigenvalues for circular Rydberg states with $m$ from $-20$ to $-60$
in units of Hartree and as a function of magnetic field strengths ranging from
$\gamma=0$ to $10^4$. The uncertainty of each eigenvalue is $\pm$ 1 in the
last digit.}
\vspace{2mm}
\begin{tabular}{cccccccccccc}
\toprule
\multicolumn{1}{c}{$\gamma$} &&
    \multicolumn{1}{c}{$m$ = $-$ 20} &&
    \multicolumn{1}{c}{$m$ = $-$ 30} &&
    \multicolumn{1}{c}{$m$ = $-$ 40} &&
    \multicolumn{1}{c}{$m$ = $-$ 50} &&
    \multicolumn{1}{c}{$m$ = $-$ 60} \\
\colrule
0     &&$-$1.133 786 848[$-$3]$^a$&&$-$5.202 913 632[$-$4]&&$-$2.974 419 988[$-$4]&&$-$1.922 337 562[$-$4]&&$-$1.343 724 805[$-$4]\\
1.0[$-$6]&&$-$1.144 261 381[$-$3]&&$-$5.356 722 275[$-$4]&&$-$3.175 806 154[$-$4]&&$-$2.168 754 327[$-$4]&&$-$1.631 359 801[$-$4]\\
1.0[$-$5]&&$-$1.236 246 089[$-$3]&&$-$6.636 401 779[$-$4]&&$-$4.696 519 268[$-$4]&&$-$3.803 017 620[$-$4]&&$-$3.289 205 886[$-$4]\\
1.0[$-$4]&&$-$1.966 965 267[$-$3]&&$-$1.422 940 899[$-$3]&&$-$1.180 188 615[$-$3]&&$-$1.033 891 190[$-$3]&&$-$9.327 609 136[$-$4]\\
1.0[$-$3]&&$-$5.103 520 227[$-$3]&&$-$4.070 784 451[$-$3]&&$-$3.495 816 062[$-$3]&&$-$3.114 743 340[$-$3]&&$-$2.837 744 879[$-$3]\\
1.0[$-$2]&&$-$1.503 990 327[$-$2]&&$-$1.228 139 915[$-$2]&&$-$1.065 234 252[$-$2]&&$-$9.543 296 504[$-$3]&&$-$8.724 853 409[$-$3]\\
1.0[$-$1]&&$-$4.491 254 575[$-$2]&&$-$3.703 391 203[$-$2]&&$-$3.228 379 726[$-$2]&&$-$2.901 538 663[$-$2]&&$-$2.658 741 866[$-$2]\\
1.0[+0]   &&$-$1.313 427 615[$-$1]&&$-$1.091 732 328[$-$1]&&$-$9.563 906 898[$-$2]&&$-$8.625 592 784[$-$2]&&$-$7.924 859 065[$-$2]\\
1.0[+1]   &&$-$3.679 986 856[$-$1]&&$-$3.089 274 257[$-$1]&&$-$2.724 004 972[$-$1]&&$-$2.468 578 160[$-$1]&&$-$2.276 596 813[$-$1]\\
1.0[+2]   &&$-$9.671 875 648[$-$1]&&$-$8.226 052 672[$-$1]&&$-$7.317 911 994[$-$1]&&$-$6.675 852 687[$-$1]&&$-$6.189 199 560[$-$1]\\
1.0[+3]   &&$-$2.336 448 829[+0]&&$-$2.019 554 474[+0]&&$-$1.816 800 033[+0] &&$-$1.671 536 222[+0]&&$-$1.560 288 457[+0]\\
1.0[+4]   &&$-$5.109 318 515[+0]&&$-$4.496 888 304[+0] &&$-$4.097 309 019[+0] &&$-$3.806 887 190[+0]&&$-$3.581 931 029[+0]\\
\botrule
\multicolumn{6}{l}{$^aA[B] = A \times 10^{B}$}
\end{tabular}
\end{center}
\label{circularII}
\end{sidewaystable*}


\pagebreak
\centerline{Figure Captions}
\begin{description}
\item[Fig.1] (Color online) Contour plots of probability densities of the electron
in a circular state with $m=-$24, integrated over the angular variable of
cylindrical coordinates ($\rho, z, \phi$). The magnetic field is along the
$z$-axis. The total wave function is normalized such that
$\int|\Psi(\rho, z, \phi)|^2\rho d\rho dz d\phi$=1. Note: The data aspect
ratio is taken to be 1:1 between $z$ and $\rho$.
\item[Fig.2] (Color online) Probability densities of the electron in a non-circular
state $7g_{-4}$, integrated over the angular variable of cylindrical coordinates
($\rho, z, \phi$). The direction of magnetic fields and the normalization of
wave function are the same as those in Fig.1.
\end{description}

\end{document}